\documentclass[]{aa} 
\usepackage{graphicx}
\usepackage{longtable}
\begin{document}
\def\ffam {\hbox{$\,.\!\!^{\prime}$}}
\def\ffas {\hbox{$\,.\!\!^{\prime\prime}$}}
\def\ffM {\hbox{$\,.\!\!^{\rm M}$}}
\def\ffm {\hbox{$\,.\!\!^{\rm m}$}}
\def\ffs {\hbox{$\,.\!\!^{\rm s}$}}
\newcommand\nodata{ ~$\cdots$~ }
   
\title{Extragalactic H$_2$O masers and X-ray absorbing column densities
       \thanks{Based on observations with the 100-m telescope of the MPIfR 
        (Max-Planck-Institut f{\"u}r Radioastronomie) at Effelsberg} 
       \thanks{M. Kadler, present address: Exploration of the Universe Division,
        Code 662, NASA/Goddard Space Flight Center, 4800 Greenbelt Road,
        Greenbelt, MD 20771, USA}}

%\subtitle{ }

\author{J. S. Zhang\inst{1,2,3}
        \and
        C. Henkel\inst{1}
        \and
        M. Kadler\inst{1}
        \and
        L. J. Greenhill\inst{4}
        \and
        N. Nagar\inst{5}
        \and
        A. S. Wilson\inst{6}
        \and
        J. A. Braatz\inst{7}}

\offprints{J. S. Zhang
\email{jzhang@mpifr-bonn.mpg.de}}

\institute{Max-Planck-Institut f\"ur Radioastronomie, Auf dem H\"ugel 69, D-53121 Bonn, Germany
           \and
           Purple Mountain Observatory, Chinese Academy of Sciences, 210008 Nanjing, China
           \and
           Center For Astrophysics, GuangZhou University, GuangZhou, 510400, China
           \and
           Harvard-Smithsonian Center for Astrophsics, 60 Garden Street, Cambridge, MA 02138, USA
           \and
           Universidad de Concepci{\'o}n, Grupo de Astronom\'{\i}a, Casilla 4009, Concepci{\'o}n, Chile
           \and
           University of Maryland, Dept. of Astronomy, College Park, MD 20742, USA
           \and
           National Radio Astronomy Observatory, 520 Edgemont Rd., Charlottesville, VA 22903, USA}

\date{2 September 2005 / 5 December 2005}

\abstract{Having conducted a search for the $\lambda$$\sim$1.3\,cm (22\,GHz) water vapor line towards galaxies with nuclear 
activity, large nuclear column densities or high infrared luminosities, we present H$_2$O spectra for NGC\,2273, UGC\,5101, 
and NGC\,3393 with isotropic luminosities of 7, 1500, and 400\,L$_{\odot}$. The H$_2$O maser in UGC\,5101 is by far the most 
luminous yet found in an ultraluminous infrared galaxy. NGC\,3393 reveals the classic spectrum of a `disk maser', represented 
by three distinct groups of Doppler components. As in all other known cases except NGC\,4258, the rotation velocity of the 
putative masing disk is well below 1000\,km\,s$^{-1}$. Based on the literature and archive data, X-ray absorbing column densities 
are compiled for the 64 galaxies with reported maser sources beyond the Magellanic Clouds. For NGC\,2782 and NGC\,5728, we present 
{\it Chandra} archive data that indicate the presence of an active galactic nucleus in both galaxies. Modeling the hard 
nuclear X-ray emission, NGC\,2782 is best fit by a high energy reflection spectrum with $N_{\rm H}$$\ga$10$^{24}$\,cm$^{-2}$. 
For NGC\,5728, partial absorption with a power law spectrum indicates $N_{\rm H}$$\sim$8$\times$10$^{23}$\,cm$^{-2}$. The 
correlation between absorbing column and H$_2$O emission is analyzed. There is a striking difference between kilo- and 
megamasers with megamasers being associated with higher column densities. All kilomasers ($L_{\rm H_2O}$$<$10\,L$_{\odot}$) 
except NGC\,2273 and NGC\,5194 are Compton-thin, i.e. their absorbing columns are $<$10$^{24}$cm$^{-2}$. Among the H$_{2}$O 
megamasers, 50\% arise from Compton-thick and 85\% from heavily obscured ($>$10$^{23}$cm$^{-2}$) active galactic nuclei. These 
values are not larger but consistent with those from samples of Seyfert 2 galaxies not selected on the basis of maser emission. 
The similarity in column densities can be explained by small deviations in position between maser spots and nuclear X-ray source 
and a high degree of clumpiness in the circumnuclear interstellar medium.
}

\titlerunning{Masers and column densities}

\authorrunning{J. S. Zhang}

\maketitle

%________________________________________________________________

\section{Introduction}

The $J_{\rm K_aK_c} = 6_{16}-5_{23}$ line of ortho-H$_2$O at 22\,GHz ($\lambda$$\sim$1.3\,cm) is one of the strongest and 
most remarkable spectral features of the electromagnetic spectrum. The transition, connecting levels located approximately 
645\,K above the ground state, traces warm ($T_{\rm kin}$$\ga$400\,K) and dense ($n$(H$_2)$$\ga$10$^{7}$\,cm$^{-3}$) 
molecular gas (e.g. Kylafis \& Norman 1987, 1991). Observed as a maser, the line can reach isotropic luminosities in 
excess of 10$^4$\,L$_{\odot}$ (Barvainis \& Antonucci 2005). Being traditionally known to trace oxygen-rich red giant 
stars and to pinpoint sites of (massive) star formation, it has more recently become a major tool to determine geometric 
distances and three dimensional motions of nearby galaxies and to elucidate the nuclear environment of active galaxies, 
allowing us to map accretion disks and to determine masses of nuclear engines (for recent reviews, see Greenhill 2002, 
2004; Maloney 2002; Henkel \& Braatz 2003; Morganti et al. 2004; Henkel et al. 2005a; Lo 2005; for 3-D motions, see 
Brunthaler et al. 2005). 

Extragalactic H$_2$O masers have been observed for a quarter of a century (Churchwell et al. 1977) and emission has been 
reported from more than 60 galaxies (e.g., Henkel et al. 2005b, Kondratko et al. 2006). The masers can be classified 
as related (1) to star formation, (2) to nuclear accretion disks (`disk-masers'), (3) to interactions of nuclear jet(s) 
with ambient molecular clouds or to amplification of the jet's seed photons by suitably located foreground clouds (`jet-masers') 
and (4) to nuclear outflows. H$_2$O masers with apparent (isotropic) luminosities $L_{\rm H_2O}$ $<$ 10\,L$_{\odot}$, 
often associated with sites of massive star formation, are referred to as `kilomasers', while the stronger sources, known 
to arise from the innermost parsecs of their parent galaxy, are referred to as `megamasers'. The statistical properties 
of the sample of extragalactic H$_2$O masers are not well explored. The first and last such comprehensive study, 
that of Braatz et al. (1997), dates back to when little more than a dozen sources were known. 

In order to improve our knowledge of the statistical properties of the greatly enlarged maser sample, we analyze 
the correlation between maser emission and X-ray absorbing column density toward the nuclear engine. In Sects.\,2--3 we 
summarize recent observations and show spectra of sources detected in this work. In Sect.\,4, we present X-ray 
column densities for the entire extragalactic maser sample, including a detailed analysis of {\it Chandra} archive data 
of the megamaser galaxies NGC\,2782 and NGC\,5728. Specific statistical properties of the sample involving X-ray absorbing 
columns and maser luminosities are discussed in Sect.\,5. Sect.\,6 summarizes the main results.

\section{Observations}

Our measurements were obtained with the 100-m telescope of the MPIfR at Effelsberg in June 2001, October 2004 and 
January and February 2005. The full beam width to half power was 40$''$. A dual channel HEMT receiver provided 
system temperatures of 100--350\,K on a main beam brightness temperature scale. The observations were carried out 
in a dual beam switching mode with a beam throw of 2$'$ and a switching frequency of $\sim$1\,Hz. The autocorrelator 
backend was split into eight bands of width 40 or 80\,MHz and 512 or 256 channels. Each band could be shifted 
individually by up to $\pm$250\,MHz ($\pm$3000\,km\,s$^{-1}$) relative to the recessional velocity of the galaxy. 
Channel spacings were $\sim$1 and $\sim$4\,km\,s$^{-1}$. 

The pointing accuracy was better than 10$''$. Calibration was obtained by measuring NGC\,7027, 3C286, or W3(OH) (for 
flux densities, see Baars et al. 1977; Ott et al. 1994). The data were reduced using the CLASS and GREG packages of 
the GILDAS software.

\begin{table*}
\caption[]{H$_2$O maser observations
\label{rms}}
\begin{scriptsize}
\begin{center}
\begin{tabular}{l c c r c c r @{--} c r c c l @{\ } l}
\hline
Source              & R.A.       & Dec.             & $V_{\rm sys}$    & Type$^{\rm a)}$      & $L_{\rm FIR}^{\rm b)}$     &  
\multicolumn{2}{c}{$V$--range}   & rms    & Channel & Epoch$^{\rm c)}$ & \multicolumn{2}{c}{$N_{\rm H}^{\rm d)}$}    \\
                    &            &                  &                  &                      & (log)                      &
\multicolumn{2}{c}{\ }           &                  &    Width         &                      &                      \\
                    &\multicolumn{2}{c}{($J$\,2000)}&   (c$z$)         &                      &                            &
\multicolumn{2}{c}{(km\,s$^{-1}$)}&  (mJy)          & (km\,s$^{-1}$)   & & \multicolumn{2}{c}{($10^{23}$cm$^{-2}$)}  \\
\hline 
\\
NGC\,1667           &04 48 37.1& $-$06 19 12 & 4547 &Sy 2  &  11.0 &   4000&5100 & 10 &1.1& 0205 & $\ge10$            & B; B99    \\
NGC\,2273           &06 50 08.7&   +60 50 45 & 1840 &Sy 2  &  10.1 &   1400&2400 &  9 &1.1& 0105 & $\geq 18$          & X; Gua05  \\
IRAS\,07145--2914   &07 16 31.2& $-$29 19 29 & 1698 &Sy 2  &  10.0 &   1100&2200 & 77 &1.1& 1004 & $\ge100$           & B; B99    \\
                    &          &             &      &      &       &   1050&2200 & 35 &1.1& 0105 &                    &           \\
UGC\,5101           &09 35 51.6&   +61 21 11 &11809 &ULIRG &  11.9 &  10550&12800&  2 &4.2& 1004 & $13\pm 2$          & C; Arm04  \\ 
NGC\,2992$^{\rm e)}$&09 45 42.0& $-$14 19 35 & 2311 &Sy 1.9&\nodata&   1770&2900 & 11 &1.1& 0205 & $\sim0.1$          & B; Gua05  \\
NGC\,3393           &10 48 23.4& $-$25 09 43 & 3750 &Sy 2  &  10.4 &   2880&4800 & 14 &4.2& 0105 & $44^{+25}_{-11}$   & X; Gua05  \\
IRAS\,11058--1131   &11 08 20.3& $-$11 48 12 &16437 &Sy 2  &  11.2 &  15500&17500&  7 &4.2& 0105 & $>100$             & B; Ris00  \\
NGC\,4418           &12 26 54.6& $-$00 52 39 & 2179 &Sy 2  &  10.9 &   1550&2900 & 33 &1.1& 1004 & $>10$              & C; Mai03  \\
Mrk\,231            &12 56 14.2&   +56 52 25 &12642 &ULIRG &  12.2 &  12050&13200&  7 &1.1& 0205 & $\sim20$           & X,B; Bra04\\
NGC\,4939$^{\rm e)}$&13 04 14.4& $-$10 20 23 & 3110 &Sy 2  &  10.3 &   2500&3770 &  8 &1.1& 0105 & $1.5^{+0.4}_{-0.5}$& X; Gua05  \\
                    &          &             &      &      &       &   2570&3700 & 11 &1.1& 0205 &                    &           \\
NGC\,4968           &13 07 06.0& $-$23 40 37 & 2957 &Sy 2  &  10.2 &   2300&3660 & 30 &1.1& 1004 & $\ge10$            & B; B99    \\
NGC\,5005$^{\rm e)}$&13 10 56.2&   +37 03 33 &  946 &LINER &  10.2 &    400&1550 & 10 &1.1& 0205 & $>10$              & B; Ris99  \\
                    &          &             &      &      &       &\multicolumn{2}{c}{\ }& & &  & $0.3^{+0.2}_{-0.2}$& X; Gua05  \\
Mrk\,668            &14 07 00.4&   +28 27 15 &22957 &RL    &\nodata&  22340&23630&  9 &1.1& 1004 & $>100$             & X; Gua04  \\
Arp\,220$^{\rm e)}$ &15 34 57.1&   +23 30 11 & 5434 &ULIRG &  12.1 &   5250&5720 & 13 &1.1& 0601 & 0.3                & C; Cle02  \\ 
NGC\,6552           &18 00 07.3&   +66 36 54 & 7942 &Sy 2  &  11.0 &   7400&8500 &  9 &1.1& 1004 & $6\pm1$            & A; B99    \\
                    &          &             &      &      &       &   7400&8500 &  8 &1.1& 0205 &                    &           \\
TXS\,1946+708       &19 45 53.5&   +70 55 49 &30228 &RL    &\nodata&  29000&31650&  5 &4.2& 1004 & $>28$              & B; Ris03  \\
IRAS\,20210+1121    &20 23 25.4&   +11 31 35 &16905 &Sy 2  &  11.7 &  16650&17200&  8 &1.1& 0105 & $\le0.6$           & B; Delu04 \\
NGC\,7674           &23 27 56.7&   +08 46 45 & 8671 &Sy 2  &  11.5 &   8100&9380 & 19 &1.1& 0105 & $\ge100$           & B; B99    \\
\\
\hline
\end{tabular}
\end{center}

a) LINER = Low Ionization Nuclear Emission Line Region; RL = Radio Loud; Sy = Seyfert; ULIRG = UltraLuminous InfraRed Galaxy \\
b) Infrared flux densities taken from the IRAS Point Source Catalog (Fullmer \& Lonsdale 1989) or from NED were converted into 
   far infrared luminosities following the procedure outlined by Wouterloot \& Walmsley (1986); $H_0$=75\,km\,s$^{-1}$\,Mpc$^{-1}$; 
   the values given are the logarithms of the infrared luminosity in solar units \\
c) Month (first two digits) and year (last two digits) \\ 
d) Notes concerning X-ray absorbing hydrogen column densities: 
   X-ray telescope: A: {\it ASCA}; B: {\it BeppoSAX}; C: {\it Chandra}; X: {\it XMM-Newton};
   References: Arm04: Armus et al. (2004); B99: Bassani et al. (1999); Bra04: Braito et al. (2004); Cle02: Clements et al. (2002); 
   Delu04: Deluit (2004); Gua04: Guainazzi et al. (2004a); Gua05: Guainazzi et al. (2005a); Mai03: Maiolino et al. (2003);
   Ris99: Risaliti et al. (1999); Ris00: Risaliti et al. (2000); Ris03: Risaliti et al. (2003) \\
e) NGC\,2992: a revived AGN according to Gilli et al. (2000);
   NGC\,4939: a source varying from Compton-thick to Compton-thin, see Gua05;
   NGC\,5005: a possibly missclassified Compton-thin source from Gua05;
   Arp\,220: the column density from Cle02 refers to the western nucleus and is highly uncertain. 
\end{scriptsize}
\end{table*}

\section{Results}

\begin{figure}[ht]
\vspace{-7.6cm}
\hspace{0.9cm}
\includegraphics[bb=58 39 548 571, angle=-90, width=15.0cm]{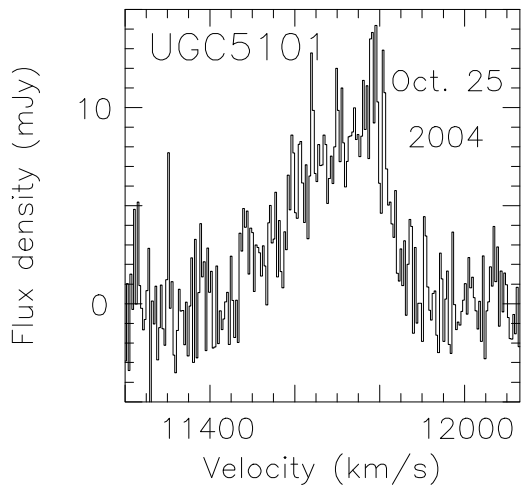}
\vspace{0.4cm}
\caption[fig1]{The $\lambda$$\sim$1.3\,cm (22\,GHz) H$_2$O megamaser profile toward UGC\,5101, with a channel
spacing of 4.2\,km\,s$^{-1}$. $\alpha_{2000}$ = 09$^{\rm h}$ 35$^{\rm m}$ 51\ffs6, $\delta_{2000}$ = 61$^{\circ}$ 
21\arcmin\ 11\arcsec. Velocity scales are with respect to the Local Standard of Rest (LSR) and use the optical convention 
that is equivalent to c$z$. $V_{\rm sys}$ = c$z_{\rm sys}$ = 11809\,km\,s$^{-1}$ (NASA/IPAC Extragalactic Database 
(NED)). $V_{\rm LSR}$--$V_{\rm HEL}$ = +3.69\,km\,s$^{-1}$. 
\label{ugc5101}}
\end{figure}

\begin{figure}[ht]
\vspace{-7.6cm}
\hspace{0.9cm}
\includegraphics[bb=58 39 548 571, angle=-90, width=15cm]{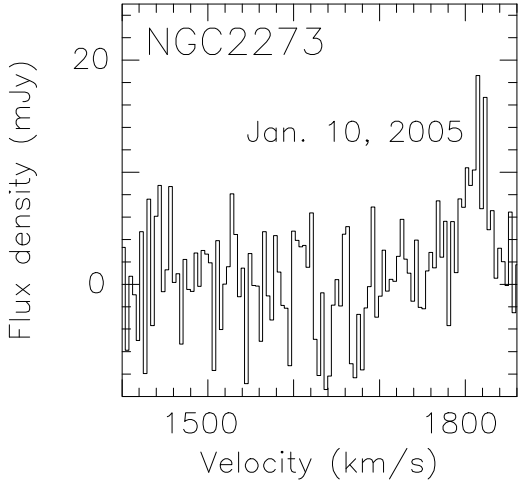}
\vspace{0.4cm}
\caption[fig1]{22\,GHz H$_2$O maser profile smoothed to a channel spacing of 4.2\,km\,s$^{-1}$ toward NGC\,2273. No emission 
was seen between 1860 and 2400\,km\,s$^{-1}$ (see Table~\ref{rms}). $\alpha_{2000}$ = 06$^{\rm h}$ 50$^{\rm m}$ 08\ffs7, 
$\delta_{2000}$ = 60$^{\circ}$ 50\arcmin\ 45\arcsec. $V_{\rm sys}$ = c$z_{\rm sys}$ = 1840\,km\,s$^{-1}$ (NED). 
$V_{\rm LSR}$--$V_{\rm HEL}$ = +0.46\,km\,s$^{-1}$.
\label{ngc2273}}
\end{figure}

Table~\ref{rms} summarizes observations of our sample of galaxies showing nuclear activity, high nuclear column densities (see 
Sect.\,4) or high infrared luminosities. Among the 18 sources observed, three were detected. Figs.\,\ref{ugc5101}--\ref{ngc3393} 
show the line profiles of the detected targets. Table~\ref{gauss} provides line parameters obtained from Gaussian fits. Adopting 
a Hubble constant of $H_{0}$ = 75\,km\,s$^{-1}$\,Mpc$^{-1}$, to be consistent with previous maser studies and to be close to the 
parameters of the standard $\Lambda$-cosmology (e.g. Spergel et al. 2003), isotropic luminosities are $L_{\rm H_2O}$ $\sim$ 7, 
1500, and 400\,L$_{\odot}$ for NGC\,2273, UGC\,5101 and NGC\,3393, respectively.

\subsection{UGC\,5101}

UGC\,5101 hosts one of the most luminous megamasers yet detected. Along with NGC\,6240 (Hagiwara et al. 2002; Nakai et al. 
2002; Braatz et al. 2003), UGC\,5101 contains only the second H$_2$O maser so far encountered in an ultraluminous infrared 
galaxy (ULIRG). Interestingly, neither of the two ULIRGs with H$_2$O maser emission is known to host an OH megamaser. 

As indicated in Sect.\,1, all H$_2$O megamasers studied with milliarcsecond resolution have been found to be associated 
with the active galactic nucleus (AGN) of their parent galaxy. Is this also the case for ULIRGs, which are characterized by 
extremely high rates of massive star formation? While masers associated with star formation are not known to reach luminosities 
in excess of 10\,L$_{\odot}$ (for the highest luminosities reached, see Tarchi et al. 2002b), the superposition of a large 
number of such sources {\it might} lead to cumulative luminosities that rival those of luminous nuclear megamasers. There 
is, however, evidence against this case. A large number of masers with luminosities of 0.01--1\,L$_{\odot}$ should lead to 
detectable H$_2$O emission not only in UGC\,5101, but also in Arp\,220. Because of the large number of such putative star 
forming H$_2$O maser sources, their overall luminosity should be fairly well correlated with star formation rates and thus 
with infrared luminosity. These luminosities are quite similar ($\sim$10$^{12}$\,L$_{\odot}$) for most ULIRGs. Arp\,220, 
which has slightly less than half the redshift of UGC\,5101, should thus show $\sim$5 times stronger H$_2$O emission. 
However, Arp\,220 was not detected (see Table 1), thus hinting at a nuclear origin of the maser in UGC\,5101.

It is known that UGC\,5101 hosts a prominent buried AGN. Its LINER nucleus contains an absorbed hard X-ray component 
with a 6.4\,keV spectral feature (Imanishi et al. 2003). Its central radio continuum is dominated by an AGN, not by a 
starburst (Lonsdale et al. 2003), and infrared signatures also point towards a dominant nuclear engine that is highly 
obscured (Imanishi et al. 2001; Armus et al. 2004). In the other known ULIRG hosting a water vapor maser, NGC\,6240, the 
H$_2$O emission was found to spatially coincide with one of the two nuclei of the interacting pair of galaxies (Hagiwara 
et al. 2003). We thus conclude that the maser in UGC\,5101 must be nuclear.

\subsection{NGC\,2273}

NGC\,2273 is a barred, early-type spiral galaxy with a Seyfert 2 nucleus (e.g. Petitpas \& Wilson 2002) and was among 
the galaxies observed in the Braatz et al. (1996) survey. The sensitivity of their measurements was, however, not 
high enough to detect a feature at the 15\,mJy level. NGC\,2273 is by far the least luminous of the three detected 
maser sources presented here. With $L_{\rm H_2O}$$<$10\,L$_{\odot}$, it is a `kilomaser' and may be associated with a
prominent site of star formation instead of an AGN (e.g. Tarchi et al. 2002b; Henkel et al. 2005b). The CO emission in 
the innermost few arcsec of the galaxy (Petitpas \& Wilson 2002), the nuclear dust ring of size 14\arcsec\ (Yankulova 1999), 
and the optical emission lines (e.g. Gu et al. 2003) are all indicative of ongoing star formation that might trigger maser 
emission at the observed luminosity. On the other hand, molecular gas is apparently streaming toward the innermost regions 
of the galaxy, which hosts a Compton-thick central X-ray source (see Sect.\,5). Thus a nuclear maser is also possible. 
The linewidth, $\Delta V$$\gg$0.3\,km\,s$^{-1}$ (Table 2), suggests that the maser is longlived. Thus interferometric 
measurements may soon provide a position with subarcsec accuracy.

\begin{figure}[ht]
\vspace{-7.6cm}
\hspace{0.9cm}
\includegraphics[bb=58 39 548 571, angle=-90, width=15.0cm]{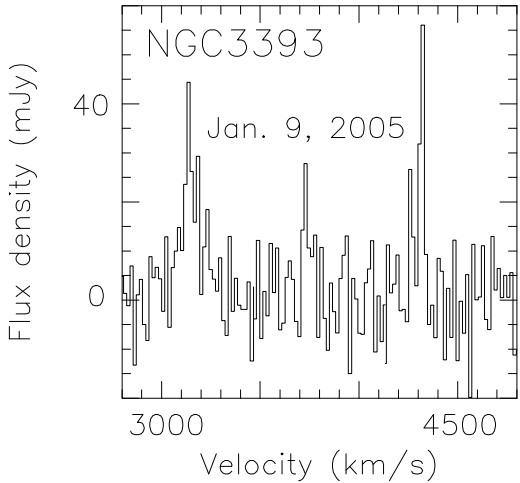}
\vspace{0.4cm}
\caption[fig1]{22\,GHz H$_2$O megamaser profile smoothed to a channel spacing of 16.8\,km\,s$^{-1}$ toward 
NGC\,3393. $\alpha_{2000}$ = 10$^{\rm h}$ 48$^{\rm m}$ 23\ffs4, $\delta_{2000}$ = --25$^{\circ}$ 09\arcmin\ 43\arcsec. 
$V_{\rm sys}$ = 3750\,km\,s$^{-1}$ (NED). $V_{\rm LSR}$--$V_{\rm HEL}$ = \hbox{--9.09\,km\,s$^{-1}$.}
\label{ngc3393}}
\end{figure}

\subsection{NGC\,3393}

Like NGC\,2273, NGC\,3393 is a barred, early-type spiral with a Seyfert 2 nucleus (e.g. Cooke et al. 2000). It is seen nearly 
face-on and was also part of the Braatz et al. (1996) survey, where it was undetected (their 1$\sigma$ noise level is 11\,mJy 
with a channel spacing of 0.66\,km\,s$^{-1}$). More recently, however, NGC\,3393 was independently detected by Kondratko et 
al. (2006). Fig.\,\ref{ngc3393} shows a lineshape reminiscent of that seen toward NGC\,4258. Thus the nuclear origin of the maser 
is evident, indicating the presence of an accretion disk that is viewed approximately edge-on and suggesting a rotation velocity 
of order 460\,km\,s$^{-1}$. It is noteworthy that none of the disk maser (candidate) sources (for NGC\,1068, NGC\,2960, NGC\,3079, 
IC\,2560, NGC\,4945, Circinus and NGC\,6323, see Gallimore et al. 2001; Henkel et al. 2002; Kondratko et al. 2005; Ishihara et 
al. 2001; Greenhill et al. 1997, 2003; Braatz et al. 2004) shows the high rotation velocity, $\sim$1000\,km\,s$^{-1}$, observed 
towards NGC\,4258, the prototypical source of its class. In view of the Toomre $Q$ parameter for disk stability (e.g. Kondratko 
et al. 2005) it seems that NGC\,4258 is hosting, with its comparatively large nuclear mass (e.g. Herrnstein et al. 1999) and 
circumnuclear rotation velocity (Nakai et al. 1993), an exceptionally compact and well ordered masing disk that is not 
typical for Seyfert galaxies.

\section{Radio versus X-ray data}

\subsection{General aspects}

Traditionally, detection rates of megamaser surveys are low (e.g. Henkel et al. 2005b). The likely most direct way 
to identify potential megamaser sources involves a determination of column densities towards the nuclear engine(s). Braatz 
et al. (1997; their Fig.\,35) proposed that megamaser galaxies may be sources with particularly large nuclear column densities. 
With only a dozen megamaser sources known at that time and with only five having measured X-ray absorbing column densities, 
however, a detailed study was not possible. 

While the presence of only one detected maser line does not strongly confine pumping scenarios, radiative transfer 
calculations provide significant constraints to kinetic temperature, H$_2$ density and  H$_2$O column density (e.g. 
Kylafis \& Norman 1991). To connect H$_2$O with total column densities along a given line-of-sight, however, fractional 
H$_2$O abundances are also needed. Such abundances were recently determined, not for distant AGN, but for some nearby 
interstellar clouds in the galactic disk (e.g. Melnick et al. 2000; Snell et al. 2000a, b, c; Neufeld et al. 2000a, 
b). The studies indicate [H$_2$O]/[H$_2$] $\sim$ 10$^{-9}$--10$^{-7}$ in cool molecular cloud cores, $\sim$10$^{-6}$ in 
stellar outflow sources and $\sim$10$^{-4}$ in warm shock heated gas. H$_2$O column densities of order 10$^{18}$ 
-- 10$^{20}$\,cm$^{-2}$ needed for efficient maser amplification (e.g. Kylafis \& Norman 1991; Neufeld et al. 1994; 
Kartje et al. 1999) then require total hydrogen column densities well in excess of 10$^{22}$\,cm$^{-2}$. 

With respect to observations of extragalactic H$_2$O masers, there is considerable progress since Braatz et al. (1997). 
The sample of known extragalactic H$_2$O masers has been quadrupled (see Henkel et al. 2005b; Kondratko et al. 2006). 
To determine column densities toward a nuclear engine requires modeling its spectrum at X-ray wavelengths. Seyfert 2 and 
LINER galaxies show nuclear X-ray sources with a sometimes heavily absorbed soft part of their spectrum. Recent advances in 
X-ray astronomy with respect to angular resolution, in order to avoid contamination by stellar sources and diffuse emission
in the vicinity of the nuclear engine, provide sensible information on column densities toward a significant number of AGN. 

Of particular interest is the group of `Compton-thick' galaxies with column densities in excess of 10$^{24}$~cm$^{-2}$. 
The bulk of this column density must arise from a circumnuclear shell or torus, since large scale ($\sim$200\,pc) column 
densities in galaxies with `normal' nuclear starbursts tend to be one to two orders of magnitude smaller (e.g. Harrison 
et al. 1999; Mao et al. 2000; Wang et al. 2004), while even in ULIRGs the corresponding values are $<$10$^{24}$\,cm$^{-2}$ 
(e.g. Pappa et al. 2000). A large column density, of order 10$^{24}$~cm$^{-2}$ or more in a highly confined region, 
presumably encompassing a few pc or less (e.g. Miyoshi et al. 1995; Gallimore et al. 2001), suggests that at least some 
fraction of the circumnuclear environment is dense and predominantly molecular. Thus a statistical comparison between 
the occurrence and strength of nuclear maser emission and column density is mandatory.

\begin{table}
\caption[]{H$_2$O line parameters obtained from Gaussian fits
\label{gauss}}
\begin{scriptsize}
\begin{center}
\begin{tabular}{c c r @{$\pm$} l r @{$\pm$} l}
\hline
Source       & $\int{S\,{\rm d}V}$  & \multicolumn{2}{c}{$V$} & \multicolumn{2}{c}{$\Delta V_{1/2}$} \ \ \\
             & (Jy\,km\,s$^{-1}$)   & \multicolumn{4}{c}{(km\,s$^{-1}$)}         \\
\hline
NGC\,2273    & 0.49$\pm$0.09        &  1813 & 3     &  34 & 9        \\
UGC\,5101    & 2.69$\pm$0.13        & 11721 & 7     & 267 & 15       \\
NGC\,3393    & 3.96$\pm$0.66        &  3151 &11     & 141 & 32       \\
             & 0.75$\pm$0.20        &  3720 & 3     &  20 & 5        \\ 
             & 0.71$\pm$0.22        &  4250 & 4     &  23 & 7        \\
             & 1.70$\pm$0.22        &  4309 & 1     &  24 & 3        \\
\hline
\end{tabular}
\end{center}
\end{scriptsize}
\end{table}

\subsection{The X-ray data}

With the 53 targets from Henkel et al. (2005b; their Table 4), the type 2 Quasar J0804+3607 (Barvainis \& Antonucci
2005), the three detections presented in Sect.\,3 and seven additional sources from Kondratko et al. (2006), we analyze 
a total of 64 H$_2$O maser galaxies beyond the Magellanic Clouds. In order to study the correlation between column 
density and maser emission, we searched for published column densities in the literature, mainly based on {\it ASCA}, 
{\it BeppoSAX}, {\it XMM-Newton} and {\it Chandra} X-ray spectroscopic observations. The data, including {\it RXTE} 
(Rossi X-ray Timing Explorer) observations, are summarized in Tables~\ref{limit}--\ref{coldensity}. 

The sample of 64 objects includes 13 kilomasers ($L_{\rm H_2O}$$<$10\,L$_{\odot}$, their source names are marked 
by italics in Table~\ref{coldensity}) and 51 megamasers, the latter presumably entirely of nuclear origin. Except 
NGC\,4293, we find absorbing columns for all the kilomasers in the literature. Beside of NGC\,5194 (M\,51) and 
NGC\,2273, all are Compton-thin. NGC\,5194 appears to be variable from marginally Compton-thick in 1993 (Terashima et al. 1998; 
Bassani et al. 1999) to Compton-thick in 2000 (Fukazawa et al. 2001) and NGC\,2273, the newly detected maser source, is a 
Compton-thick source detected by {\it BeppoSAX} (Maiolino et al. 1998) and {\it XMM-Newton} (Guainazzi et al. 2005a).

\begin{table}
\caption[]{X-ray flux limits for H$_2$O maser sources without good spectra, observed and corrected for galactic absorption 
(see Sect.\,4.2). With the exception of the kilomaser galaxy NGC\,4293, all sources host megamasers. 
\label{limit}}
\begin{scriptsize}
\begin{center}
%\begin{flushleft}
\begin{tabular}{lcccc}
\hline\noalign{\smallskip}
Source  &    Observed &   Corrected  \\
        &    upper    &   upper      \\
        &    limit    &   limit      \\
        & \multicolumn{2}{c}{(10$^{-13}$\,erg\,cm$^{-2}$\,s$^{-1}$)} \\
\hline\noalign{\smallskip}

NGC\,235A$^{\rm a)}$         & 2.91 &  4.50  \\
NGC\,449$^{\rm b)}$          & 3.07 &  6.55  \\
NGC\,591$^{\rm b)}$          & 3.73 &  7.63  \\
NGC\,613$^{\rm a)}$          & 2.14 &  3.38  \\
IC\,184$^{\rm a)}$     & 2.51 &  4.19  \\
UGC\,3255$^{\rm a)}$         & 3.66 & 10.34  \\
VII\, ZW 73$^{\rm a)}$       & 4.02 &  9.35  \\
Mrk\,78$^{\rm a)}$           & 5.88 & 11.55  \\
J0804+3607$^{\rm a)}$        & 2.06 &  3.19  \\
NGC\,2824$^{\rm a)}$         & 3.02 &  5.50  \\
NGC\,2960$^{\rm a)}$         & 3.41 &  6.51  \\
NGC\,2979$^{\rm a)}$         & 2.93 &  5.75  \\
Mrk\,34$^{\rm b)}$           & 1.60 &  2.16  \\
NGC\,3735$^{\rm a)}$         & 1.64 &  2.52  \\
NGC\,4293$^{\rm a)}$         & 2.74 &  4.78  \\
NGC\,5495$^{\rm a)}$         & 4.58 &  9.22  \\
NGC5793$^{\rm b)}$           & 9.41 &  22.8  \\
NGC\,6323$^{\rm a)}$         & 1.61 &  2.67  \\
IRAS F19370--0131$^{\rm a)}$ & 5.11 & 15.00  \\
NGC\,6926$^{\rm a)}$         & 3.20 &  7.49  \\
AM2158-380NED02$^{\rm a)}$   & 2.77 &  4.20  \\
TXS\,2226-184$^{\rm b)}$     & 4.31 &  7.84  \\
IC\,1481$^{\rm b)}$          & 3.61 &  7.73  \\ 

\noalign{\smallskip}
\hline
\noalign{\smallskip} 
\end{tabular}
\end{center}

a) Upper limits to 0.1--2.4\,keV fluxes, obtained from the {\it ROSAT} X-Ray All-Sky Survey with HEASARC (High Energy 
Astrophysics Science Archive Research Center). \\
b) Tentative or negative detections by {\it ASCA} (0.3--12\,keV) from TARTARUS version 3.0 at formal 3.19, 9.34, 4.72, 
7.36, 0.12, and 2.89\,$\sigma$ levels for NGC\,449, NGC\,591 Mrk\,34, NGC\,5793, TXS\,2226-184 and IC\,1481, respectively. 
For the way the given flux limits were derived, see also footnote a). \\
\end{scriptsize}
\end{table}

Among the 51 megamaser sources, 27 have no published column densities. Sixteen of these were only observed 
by the {\it ROSAT} All-Sky Survey (see Table~\ref{limit}, footnote a, for flux limits without and with correction for 
galactic absorption). Another six sources were measured by {\it ROSAT} and {\it ASCA}, but not by {\it BeppoSAX}, {\it XMM-Newton} 
or {\it Chandra} (for limits, see Table~\ref{limit}, footnote b). Future observations of these targets by {\it XMM-Newton} 
and {\it Chandra} would be desirable. 

The remaining five sources without derived column densities (and not observed by {\it ASCA} or {\it BeppoSAX}) were 
recently measured by {\it Chandra} and/or {\it XMM-Newton} (IRAS F01063--8034, NGC\,2782, ESO\,269--G012, NGC\,4922, 
and NGC\,5728). There are no published column densities yet. For NGC\,4922, which was observed in November 2004, 
count rates are not high enough to derive a meaningful result. We obtained data for the other four sources from the 
{\it XMM-Newton} Science and {\it Chandra} Data Archives and extracted their spectra using the software packages 
{\sc SAS v.6.1.0} ({\it XMM-Newton} Science Analysis Software) and {\sc CIAO v.2.3} ({\it Chandra} Interactive 
Analysis of Observations, http://cxc.harvard.edu/ciao/). The XSPEC v11.2.0 package was then used to fit the spectra and 
to obtain column densities.  
    
Here we present the results of the X-ray data analysis of NGC\,2782 and NGC\,5728. The other two sources will be 
discussed elsewhere (Greenhill et al., in preparation).

\begin{figure*}[t]
\includegraphics[width=\textwidth]{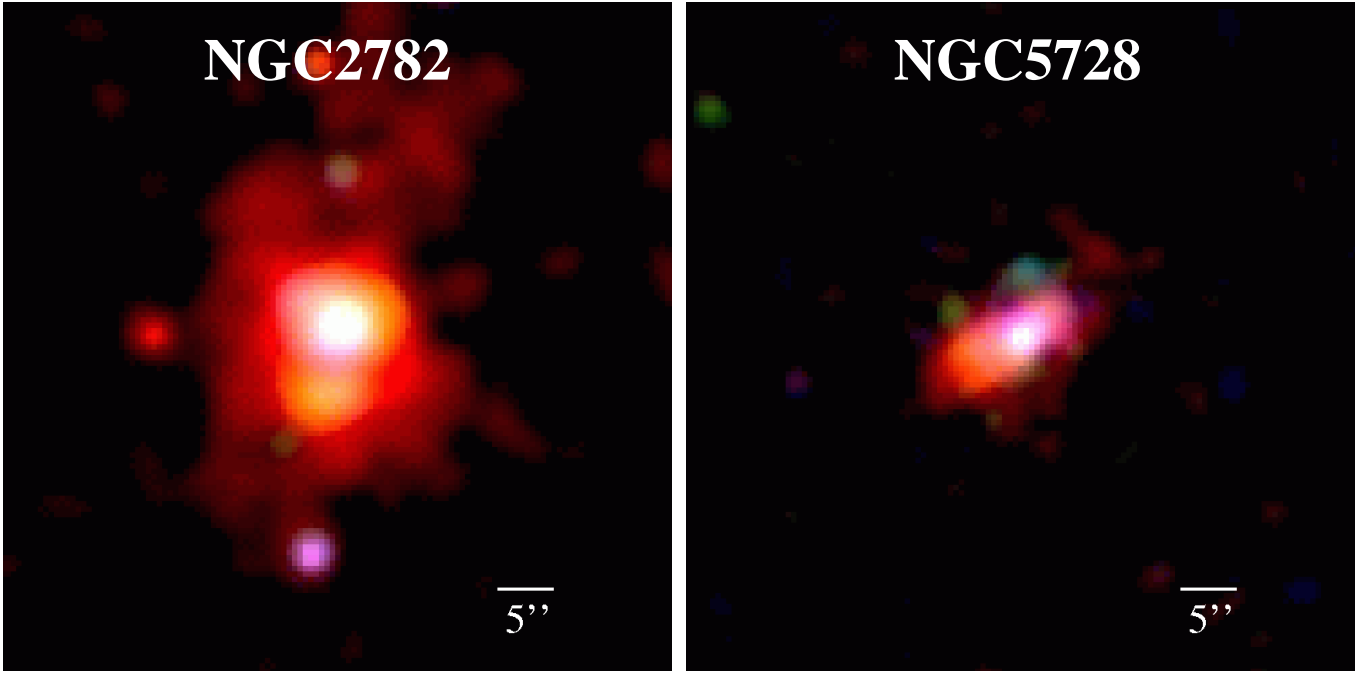}
\caption{Adaptively smoothed three-color coded {\it Chandra} images of NGC\,2782 (left) and NGC\,5728 (right). The maximal 
smoothing scales are 3$^{\prime\prime}$ and 2$^{\prime\prime}$, respectively. Red represents soft emission between 
0.3\,keV and 1.5\,keV, green corresponds to 1.5--2.5\,keV and blue denotes the energy range 2.5--8.0\,keV.}
\label{fig:ngc2782_and_ngc5728}
\end{figure*}

\subsubsection{Data acquisition and reduction}

{\it Chandra} observed NGC\,2782 on May 17, 2002\footnote{Obs. ID: 3014; PI: Ian Stevens} and NGC\,5728 on June 27, 
2003\footnote{Obs. ID: 4077; PI:  Steven Kraemer} with effective exposure times of 29.6\,ksec and 18.7\,ksec, 
respectively. Both observations were carried out with the ACIS-S detector with a standard frame time of 3.2\,s. The 
appropriate calibration database was provided by the {\it Chandra} X-ray Center and standard screening criteria were used. We 
checked that pile-up is not important due to comparatively low count rates of 0.03\,counts/frame and 0.06\,counts/frame, 
respectively, and that no flaring-background events occurred during the observations. Extended emission on scales of 
several arcseconds was found in both sources. Therefore, we extracted source spectra from circles with a radius of 
10$^{\prime\prime}$ around the nuclear positions and background spectra from nearby regions free of X-ray sources. 
Nuclear spectra from smaller circles of 1$^{\prime\prime}$ in radius were also extracted. Appropriate {\sc rmf-} and 
{\sc arf-}files were created for each individual spectrum by applying standard procedures. The energy range 0.3--8.0\,keV 
was considered. The spectra were binned to contain 25 counts per bin (15 counts per bin for the nuclear spectra).

\subsubsection{Spatial analysis}

In Fig.~\ref{fig:ngc2782_and_ngc5728}, we show adaptively smoothed, three-color {\it Chandra} images of NGC\,2782 and 
NGC\,5728. Both sources exhibit a bright, unresolved core and extended regions with emission several arcseconds in size. 
Both large-scale emitting regions have optical and radio counterparts. For NGC\,2782, the relatively weak large-scale 
emitting region is more elongated to the north than to the south, consistent with results from {\it ROSAT} (Jogee et al. 
1998). The teardrop-shaped X-ray emitting region around the nucleus of NGC\,2782 resembles the ${\rm H}\alpha$ structure 
revealed by Hubble Space Telescope (HST) imaging (Fig.\,4 in Jogee et al. 1998). The H$\alpha$ + N{\sc ii} structure 
reveals the X-ray `knot' 6$\,.\!\!^{\prime\prime}$5 south of the nucleus to be fully surrounded by a shell of line 
emitting clouds (Jogee et al. 1999). This shell is associated with a starburst-driven outflow bubble. In addition to 
a 5\,GHz radio counterpart of this southern bubble, there is a second bubble of similar size on the opposite side of 
the nucleus, only visible at radio wavelengths. The detection of a hidden AGN-like nucleus from the X-ray spectrum of 
NGC\,2782 (likely revealing a 6.4\,keV iron feature in its innermost region; see Sect.\,4.2.3) and the H$_2$O maser 
emission (Braatz et al. 2004) suggest a more dominant role of the central engine than previously anticipated.

NGC\,5728 reveals a structure elongated southeast to northwest with predominantly soft X-ray emission from 
the southeast and harder emission from the northwest. This difference may be caused by absorption of soft X-rays 
towards the northwestern region, possibly due to a geometry in which the southeastern region is closer to the observer. 
This scenario is supported by optical HST emission line studies that reveal a pronounced biconical axisymmetric structure 
of the ionization region (Wilson et al. 1993). The X-ray morphology resembles in size and extent the inner parts of the 
ionization cones in NGC\,5728.

\begin{figure*}[t]
\includegraphics[width=\textwidth]{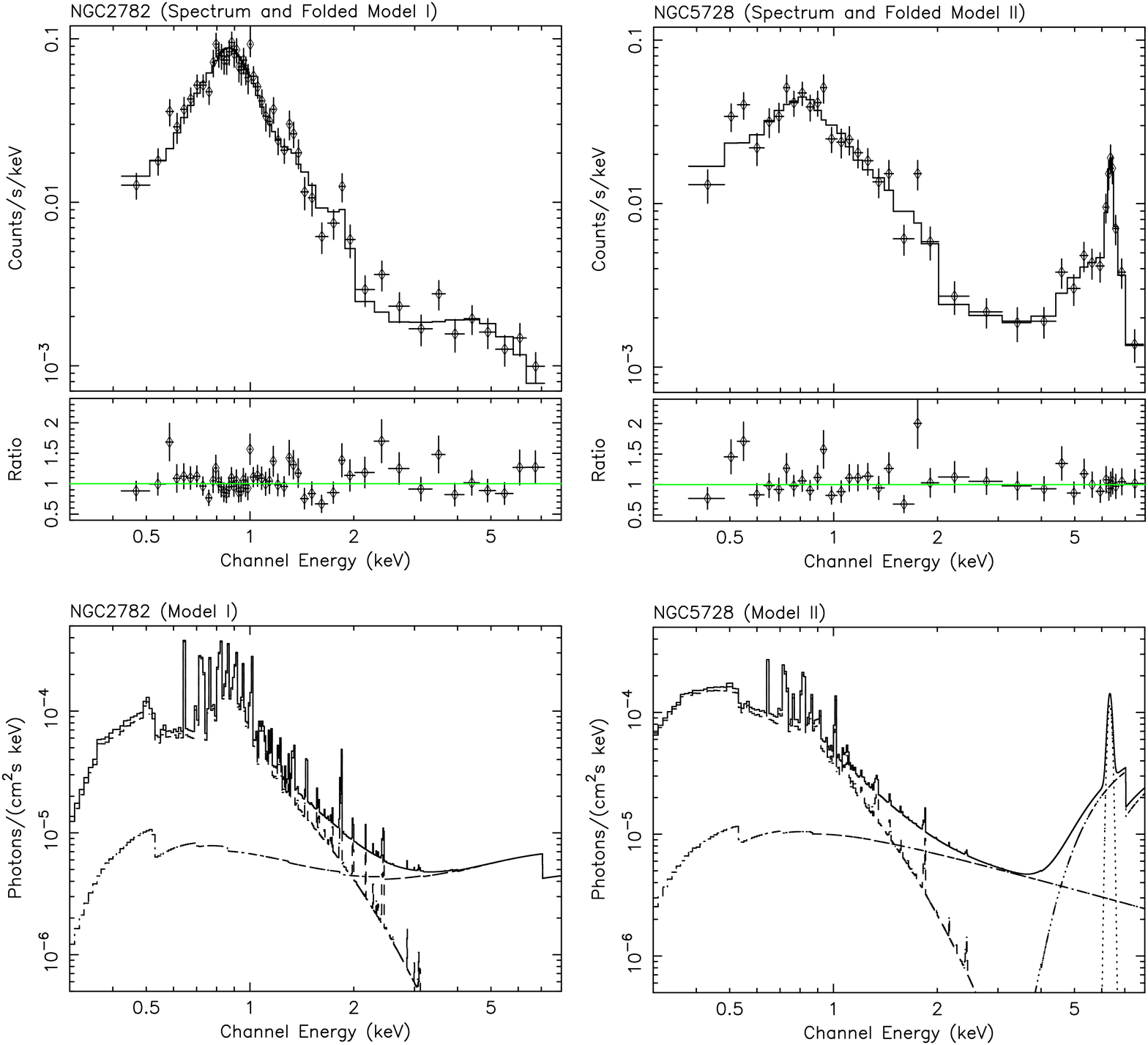}
\caption{{\it Chandra} X-ray spectra and spectral models of NGC\,2782 (left panels) and NGC\,5728 (right panels). The source 
spectra are from circles with a radius of 10$^{\prime\prime}$ around the nuclear positions. Crosses mark the observed 
fluxes and the histograms show the best-fit model, a Compton-thick high-energy reflection model for NGC\,2782 and a 
partially absorbed power law model for NGC\,5728 (upper half of the upper panels). Residuals are given below. The lower 
panels show the unfolded model spectra with their subcomponents. In the case of NGC\,2782, these are a thermal plasma 
emission ({\sc MEKAL}) and a Compton-thick reflection component, modified by absorption from the Galaxy and the redshifted 
target. In the case of NGC\,5728, the fit consists of a thermal plasma ({\sc MEKAL}), a partially absorbed power law 
component and a Gaussian shaped Fe line, modified by the absorption from the Galaxy and the target. For details, 
see Sect.\,4.2.3.} 
\label{fig:spectra}
\end{figure*}

\subsubsection{Spectral analysis}

Starting with the `canonical' Seyfert\,2 spectrum (absorbed power law plus soft excess; see e.g. Moran et al. 2001), we 
tried to reproduce the spectral shape of both sources with models of increasing complexity. It was not possible to achieve 
an acceptable fit with a simple absorbed two-power-law model in either case\footnote{In all our models we consider 
two globally absorbing components, one being representative of material inside the Galaxy, the other one for gas inside 
the measured target source.}. Not only were the reduced chi-squared values unacceptable (3.6 and 2.4, with 33 and 48 
degrees of freedom, respectively), but the resulting best-fitting photon indices became unphysically extreme (--2.5 and 
4.6 for NGC\,2782, 0.6 and 9.5 for NGC\,5728, respectively).

We considered two alternative physical models to account for the hard nuclear X-ray emission: {\sc (i)} a Compton-thick 
high-energy reflection spectrum and {\sc (ii)} a partially absorbed power law spectrum. Both models invoke a dense absorber 
towards the central source but differ physically, e.g. in the column density of the absorbing material. While model 
{\sc (i)} implies reflection of seed photons off dense matter in excess of $10^{24}$\,cm$^{-2}$ in a disk-like geometry, 
model {\sc (ii)} applies to a scenario in which a fraction of the source is absorbed by intervening material (e.g., 
associated with dense clouds) of typically $10^{22}$\,cm$^{-2}$ to a few times $10^{23}$\,cm$^{-2}$. Model {\sc (i)} 
has the additional benefit that it is sensitive to the orientation of the disk. In addition, we requested that a valid 
spectral model should fit the integrated source spectrum including both the hard compact nuclear spectrum and the 
spatially more extended soft emission. 

We modeled the soft excess via a thermal {\sc mekal} plasma model (Mewe et al. 1985). This choice appears natural and 
physically motivated in the light of the spatial distribution of the soft emission (see Sect.\,4.2.2 and 
Fig.\,\ref{fig:ngc2782_and_ngc5728}). The transition from the integrated source spectrum to the nuclear spectrum should 
thus be possible via modifications of the adjustable {\sc mekal} parameters inducing variations in intensity and possibly 
in temperature. 

\begin{table*}
\caption[]{Spectral fitting results (see Sect.\,4.2.3 and Fig.\,\ref{fig:ngc2782_and_ngc5728})
\label{fitting}}
\begin{scriptsize}
\begin{center}
\begin{tabular}{l c c c c c c c c l l l l }
\hline
Source$^{\rm a)}$      & M$^{\rm b)}$             &$N_{\rm H}$$^{\rm c)}$& kT$^{\rm d)}$       & $Z^{\rm e)}$            
& $R^{\rm f)}$         &$\theta_{\rm obs}^{\rm g}$& $\Gamma^{\rm h)}$    &$E_{\rm Fe}^{\rm i)}$& $\sigma_{\rm Fe}^{\rm j)}$   
&       EW$^{\rm k)}$  &  Flux$^{\rm l)}$         & $\chi^{2}/dof$                     \\
                       &                          &                      &  [keV]              &                      
&                      &                          &                      &  [keV]              & [ev]                 
&   [keV]              &                          &                                    \\
\hline 
\\
NGC2782$_{\rm Int}$    & ${\rm I}$            &  $>10$               &$0.66^{+0.03}_{-0.03}$&$0.15^{+0.01}_{-0.01}$&    0.004       &
$75.5^{+3.4}_{-4.7}$   &$1.68^{+0.09}_{-0.08}$&  --                  &     --               &      --              &        4.39    &
52/44          \\
                       & ${\rm II}$           &$12.7^{+4.2}_{-2.3}$  &$0.66^{+0.03}_{-0.03}$&$0.16^{+0.01}_{-0.01}$&    0.130       &
 --                    &$0.25^{+0.08}_{-0.06}$&  --                  &     --               &      --              &        4.80    & 
49/45          \\
NGC2782$_{\rm Nuc}$    & ${\rm I}$            &  $>10$               &$0.45^{+0.15}_{-0.12}$&        1.00          &    0.000       &
$70.4^{+5.3}_{-7.0}$   &$2.14^{+0.11}_{-0.11}$&$6.39^{+0.10}_{-0.12}$& $86^{+159}_{-86}$    &     0.99             &        1.86    & 
15/10          \\
                       & ${\rm II}$           & $4.3^{+2.5}_{-1.6}$  &$0.28^{+0.04}_{-0.05}$&$0.47^{+0.21}_{-0.20}$&    0.666       &
 --                    & $\sim0.07$           &$6.35^{+0.11}_{-0.09}$&    $<$160            &     1.32             &        2.02    &    15/10          \\
\\
\hline
NGC5728$_{\rm Int}$   & ${\rm I}$             &  $>10$               &$0.61^{+0.04}_{-0.03}$&$0.02^{+0.01}_{-0.01}$&    0.037       &
$2.6^{+44.7}_{-2.6}$  &$-0.81^{+0.07}_{-0.07}$&$6.33^{+0.33}_{-0.33}$&$120^{+47}_{-111}$    &    1.55              &       13.0     & 
45/26          \\
                      & ${\rm II}$            &$7.7^{+0.55}_{-0.51}$ &$0.49^{+0.04}_{-0.03}$&$0.03^{+0.01}_{-0.01}$&    0.033       &
 --                   &$0.77^{+0.07}_{-0.06}$ &$6.33^{+0.33}_{-0.33}$& $98^{+41}_{-33}$     &    1.13              &       13.2     & 
35/27          \\
NGC5728$_{\rm Nuc}$   & ${\rm I}$             & $>10$                &$0.61^{+0.08}_{-0.08}$&$0.04^{+0.01}_{-0.02}$&    0.005       &
$9.7^{+34.9}_{-9.7}$  &$-1.09^{+0.08}_{-0.07}$&$6.32^{+0.04}_{-0.03}$& $108^{+43}_{-37}$    &    1.29              &        9.54    &
23/15          \\
                      &${\rm II}$             &$8.2^{+0.53}_{-0.50}$&$0.15^{+0.01}_{-0.01}$&$0.14^{+0.03}_{-0.03}$& 0.011           &
 --                   &$0.94^{+0.08}_{-0.07}$ &$6.32^{+0.04}_{-0.04}$& $85^{+43}_{-39}$     &    0.89              &        9.34    &     16/16          \\
\\
\hline
\end{tabular}
\end{center}
a) source$_{\rm Int}$: Integrated spectra from circular areas with a radius of 10$^{\prime\prime}$ around the nuclear position; 
source$_{\rm Nuc}$: Nuclear spectra from smaller areas of 1$^{\prime\prime}$ in radius \\
b) M: Model {\rm I}: a Compton-thick high-energy reflection spectrum (wabs$*$zwabs(mekal+hrefl(pow+gau)) in XSPEC); Model 
{\rm II}: a partially absorbed power law spectrum (wabs$*$zwabs(mekal+pow+zwabs(pow+gau)) in XSPEC). The galactic 
absorbing column densities are 1.76$\times$10$^{20}$cm$^{-2}$ and 7.83$\times$10$^{20}$cm$^{-2}$ for NGC\,2782 and NGC\,5728,
respectively (from Dickey \& Lockman 1990) \\
c) X-ray absorbing column density in units of $10^{23}$cm$^{-2}$ \\
d) Temperature of the thermal {\sc MEKAL} plasma in keV \\
e) Metallicity of the thermal {\sc MEKAL} plasma as a fraction of solar metallicity (in the case of model {\rm I} of 
NGC\,2782$_{\rm Nuc}$, this relative metallicity is not constrained by the data and was therefore fixed to unity).  \\
f) Ratio between the directly seen continuum and the scattered/absorbed continuum \\
g) Disk inclination \\
h) Photon index \\
i) There is no obvious line in the integrated spectrum of NGC\,2782 (NGC2782$_{\rm int}$). Intrinsically present line emission
may be buried in the relatively bright extended continuum \\
j) Width of the line feature that is tentative in the case of NGC\,2782 (its strength does not significantly affect the other
parameters) \\
k) Equivalent width of the line feature, see footnote j \\
l) Flux from 0.3 to 8 keV in units of 10$^{-13}$erg\,cm$^{-2}$s$^{-1}$ (corrected for galactic absorption) \\
\end{scriptsize}
\end{table*}

Table~\ref{fitting} shows the results of spectral fitting. While statistically acceptable fits to the spectra of both 
sources can be obtained with both physical models, we prefer model {\sc (i)} in the case of NGC\,2782 and model {\sc (ii)} 
in the case of NGC\,5728 (see the discussion below). Spectra and fits are shown in Fig.\,\ref{fig:spectra}.

The photon index of NGC\,2782 in model {\sc (ii)} is extraordinarily flat ($\Gamma = 0.25$). This index could only 
be explained with a complex structure of the absorbing system, invoking a variation of column densities in individual 
absorbing clouds. While such a configuration cannot be excluded, the comparatively simpler model {\sc (i)} yields a 
satisfactory fit to the data with reasonable values for the crucial parameters. The large inclination of the reflecting 
disk is consistent with a Type 2 AGN but we note that this parameter is, in spite of the small formal 1$\sigma$ 
errors given in Table~\ref{fitting}, not well constrained by the available data. While the integrated spectrum does not 
require the addition of a narrow Gaussian to represent iron-line emission, excess emission is seen in the nuclear 
spectrum (radius: 1\arcsec ). The equivalent width of this tentative feature is $\sim$990\,eV.

In the case of NGC\,5728, model {\sc (i)} is statistically inferior to our preferred model {\sc (ii)}. Moreover, the 
high-energy-reflection model {\sc (i)} implies a photon index as inverted as --0.8. The partial-obscuration model {\sc 
(ii)}, on the other hand, yields a less extreme (although still comparatively flat) photon index of 0.8. The large 
equivalent width of the Fe line (1.5\,keV and 1.1\,keV for the two models, respectively) is notable. 

In spite of the discussed uncertainties, the analysis of the {\it Chandra} spectral data of NGC\,2782 and NGC\,5728 clearly 
demonstrates the presence of compact high-column-density absorbers in both systems. In NGC\,2782, the 
high-energy-reflection model suggests a Compton-thick absorber with a column density in excess of $10^{24}$\,cm$^{-2}$ 
(even the alternative model {\sc (ii)} still implies an extreme value of $\sim 1 \times 10^{24}$\,cm$^{-2}$). For 
NGC\,5728 the partial-obscuration model predicts $\sim 8 \times 10^{23}$\,cm$^{-2}$.

To summarize, we obtained X-ray column densities for 38 maser sources, 36 from the literature and 2 from processing archived data.
12 of the 13 kilomaser galaxies and 26 of the 45 megamaser galaxies are included. Because of the larger sample, the statistics 
are improved over the last such studies, those by Braatz et al. (1997; see Sect.\,4.1) and Madejski et al. (2006; their Table
3 containing 11 disk masers).

\section{Discussion}

Not all of the 38 maser sources with well defined X-ray spectra permit a meaningful comparison between H$_2$O maser
properties and column densities. A correlation would be most convincing if there were a strong dominant X-ray source
known to coincide both with the nuclear engine of its parent galaxy and the observed H$_2$O maser.

These optimal conditions cannot be fulfilled. X-ray images with the highest angular resolution, $\sim$1\arcsec, have a linear 
resolution of $\sim$10\,pc for the nearest sources (e.g. NGC\,253). For the most distant X-ray detected target, 3C403, this 
angular resolution corresponds to 1\,kpc and it is hard to imagine that X-ray emission from the (hopefully dominant) nuclear 
source is not contaminated by a large number of (hopefully weaker) stellar objects and an extended diffuse gas component. 

The maser location poses a similar problem. Towards some sources, the H$_2$O line is observed with submilliarcsecond 
accuracy. Even in the best of all cases, however, when H$_2$O is tracing an edge-on accretion disk, the situation is 
complex. Only the `systemic' H$_2$O velocity components (for a spectrum showing redshifted, blueshifted and `systemic' 
components, see Fig.\,\ref{ngc3393}) may arise from in front of the nuclear engine. The red- and blueshifted features originate 
instead from locations displaced by up to a few parsec (e.g. Gallimore et al. 2001; Kondratko et al. 2005) from the nucleus. 
Amplification of the nuclear radio continuum by unsaturated systemic maser components is another unknown parameter that may 
strongly affect maser luminosities that depend sensitively on inclination and warping of an approximately edge-on maser disk. 

There are further complications. For many of the sources with high column density, only lower limits of order 
$N_{\rm H}$$\sim$10$^{24-25}$\,cm$^{-2}$ have been derived (see Table\,\ref{coldensity}). There is a major 
uncertainty in the column density estimates related to the possibility that the X-ray emission is predominantly 
scattered light not passing through the main body of the obscuring torus. A significant fraction of the kilomasers 
are related to star forming regions well off the center of their parent galaxy. In IC\,342, neither the maser nor the 
dominant X-ray source are nuclear (Tarchi et al. 2002a; Bauer \& Brandt 2004), both being associated with {\it different} 
star forming regions. In NGC\,598 (M\,33) the `nuclear' X-ray continuum originates from an X-ray binary close to 
the dynamical center, well off the H$_2$O masers that are observed in the large-scale disk. There is even one H$_2$O 
detected galaxy, M\,82, that lacks a well defined nuclear engine.

\begin{figure}[ht]
\resizebox{8.8cm}{12.1cm}{\includegraphics{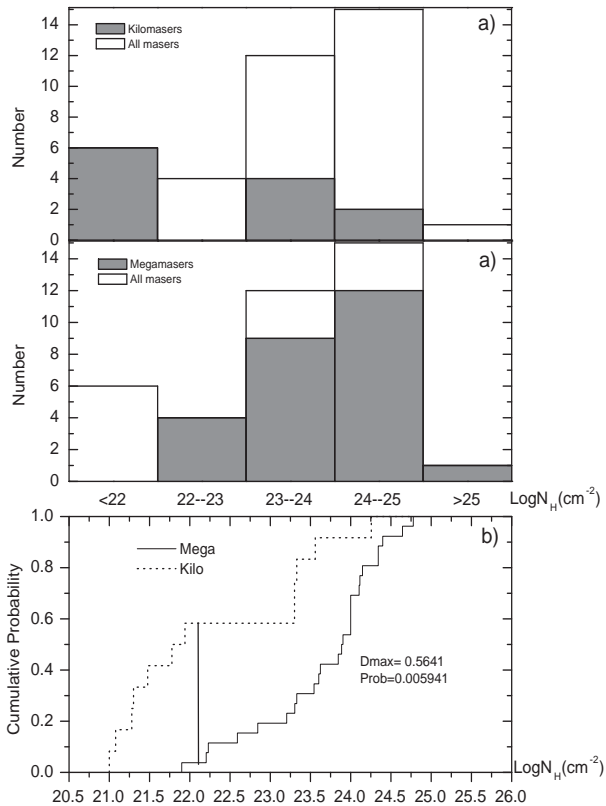}}
\caption[histo]{{\bf a}. Histograms showing the number of H$_2$O masers (kilomasers: $L_{\rm H_2O}$$<$10\,L$_{\odot}$;
megamasers: $L_{\rm H_2O}$$>$10\,L$_{\odot}$) as a function of column density determined from X-ray spectroscopy (upper panels). 
Because of angular resolution, {\it Chandra} data are taken whenever possible. Otherwise the most recent measurements are 
used (see Table~\ref{coldensity}). {\bf b}. Kolmogorov-Smirnov test for the kilomaser and megamaser samples (lower panel). 
The maximal difference of the cumulative probability (the vertical line) is 0.564 and the probability that one parent population 
would produce the kilomaser and megamaser galaxy samples is 0.006.} 
\label{histo}
\end{figure}

All these complications suggest that a correlation between H$_2$O maser emission and X-ray column density should not
only show a large scatter, but may remain entirely undetectable. For the 38 maser sources with well defined X-ray spectra, 
column densities (ignoring all the problems outlined above, preferring high resolution {\it Chandra} data whenever possible and
adopting the most recent value if {\it Chandra} data are not available) are displayed in Fig.\,\ref{histo}a. The histograms show 
the number of extragalactic H$_2$O masers as a function of column density, both for the entire sample and for the subsamples 
of 12 kilomasers and 26 megamasers. We find that there is a striking difference between the column density distribution of 
the two classes, with megamasers being associated with higher column densities.

All kilomasers (see Table\,\ref{coldensity}; their source names are printed in italics) except NGC\,2273 and NGC\,5194 are Compton-thin, 
i.e. their absorbing column densities are N$_{\rm H}$$<$10$^{24}$cm$^{-2}$. There are two groups of sources, one with column 
densities of 10$^{21-22}$\,cm$^{-2}$ and another one with column densities $>$10$^{23.3}$\,cm$^{-2}$. The mean of the 
logarithmic values is $<$log$\,N_{\rm H}$\,(cm$^{-2}$)$>$ = $22.62\pm0.37$, the error being the standard deviation of 
the mean (for NGC\,2273, the lower limit to the column density was taken; see Table\,\ref{coldensity}). 

H$_{2}$O megamasers exist preferentially in environments with higher obscuration, i.e. in circumnuclear regions of high column 
density. The peak in the distribution lies at 24 $<$log\,$N_{\rm H}$\,(cm$^{-2}$)$<$ 25. 85\% (22/26) of the megamaser sources 
are heavily obscured ($N_{\rm H}$$>$10$^{23}$cm$^{-2}$) and one half are Compton-thick ($N_{\rm H}$$\geq$10$^{24}$cm$^{-2}$). 
These fractions are consistent with the average properties of Seyfert 2 galaxies as derived by Risaliti et al. (1999) and
Guainazzi et al. (2005b), while showing slightly but not significantly higher column densities than those reported by Bassani 
et al. (1999). Taking logarithmic values, the mean column density becomes $<$log$N_{\rm H}$\,(cm$^{-2}$)$>$ = $23.77\pm0.14$. For 
six sources, there are only lower limits; these adopted limits thus lead to an average {\it below} the true mean value, likely 
{\it underestimating} the difference between megamaser and kilomaser column densities (see Table\,\ref{coldensity}). The 
Kolmogorov-Smirnov statistical method was used to test whether the distributions of column density of our kilomaser and 
megamaser sources are statistically distinguishable (see Fig.\,\ref{histo}b). The K-S test result shows that the null 
hypothesis, they both are from the same parent population, can be rejected at a confidence level of at least 99.4\%.  

In view of the uncertainties outlined above and the fact that, to date, no systematic searches for H$_2$O in galaxies with 
Compton-thick nuclear environments have been undertaken, the result is surprisingly clear: Nuclear column density is a very 
solid guide when searching for megamaser emission. Almost all megamasers known to date could have been found by analyzing 
X-ray data.

The cause of the difference between kilomaser and megamaser column densities is less clear. Differences in position between maser
and X-ray sources were already mentioned. On the one hand, we may face a high degree of small scale clumping in the observed clouds,
even within circumnuclear accretion disks. On the other hand, milliarcsecond offsets between megamaser and nuclear X-ray source are 
expected, while such offsets tend to be much larger in some of the kilomaser sources. Even accounting for the smaller mean distance
to known kilomaser galaxies, linear offsets between maser and X-ray sources tend to be much larger in kilomaser sources than for 
megamasers. The three kilomaser sources, however, that are known to be `nuclear' (i.e. the maser and dynamical centers coincide 
within 1\arcsec) and possess an associated X-ray source (NGC\,253, NGC\,4051, and NGC\,5194) reveal column densities that are 
indistinguishable from those of the megamaser sample. While three sources are clearly not sufficient for a statistically meaningful 
result, we note that this effect alone has the potential to explain the discrepancy in averaged column densities between kilomaser 
and megamaser column densities.

The subsample of (candidate) disk masers with X-ray column densities, encompassing NGC\,1068, NGC\,2639, NGC\,3079, IC\,2560, 
NGC\,3393, NGC\,4258, NGC\,4945, and Circinus, is another subsample worth to be studied. For the mean of the logarithmic values
of their column densities we find $<$log$\,N_{\rm H}$\,(cm$^{-2}$)$>$ = $23.96\pm0.36$. This value (two sources are represented 
by lower limits only) is slightly higher but still consistent with that derived for the entire megamaser sample.

\begin{figure}[ht]
\vspace{-12.0cm}
\hspace{1.3cm}
\includegraphics[bb=58 39 548 571, angle=0, width=16.9cm]{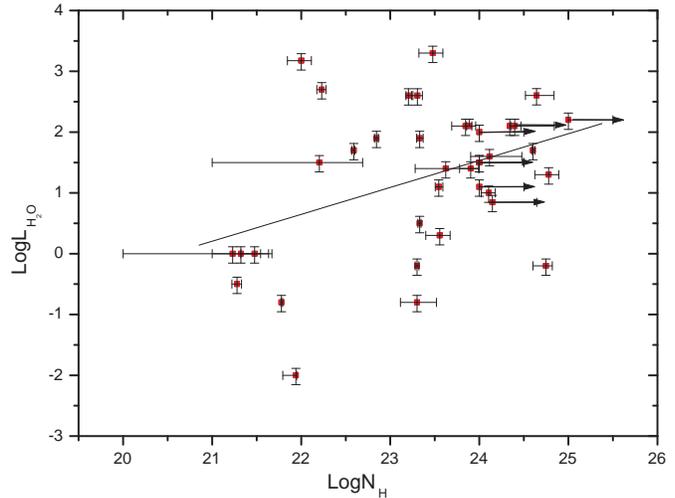}
\vspace{0.4cm}
\caption[h2oh]{Isotropic maser luminosities (in L$_{\odot}$) versus X-ray absorbing column densities $N_{\rm H}$ (in cm$^{-2}$),
including a linear fit. The data are from Table~\ref{coldensity}, taking {\it Chandra} data whenever possible and adopting otherwise 
the most recent result. The six arrows in the upper right corner represent seven sources: Mrk\,1066 and NGC\,5347 show similar
H$_2$O luminosities (Henkel et al. 2005b) and lower limits to the X-ray column density (Table\,\ref{coldensity}).  
\label{h2o}}
\end{figure}

In Fig.\,\ref{h2o}, the H$_2$O maser isotropic luminosities are plotted against their X-ray absorbing column densities. The H$_2$O 
maser isotropic luminosities were taken from Table~4 of Henkel et al. (2005b) and from Sect.\,3. We adopted a $\pm$30\% uncertainty
in maser luminosity for all sources. For the entire maser sample, again taking lower limits to the column densities as real values 
for seven targets, the linear fit gives log\,$L_{\rm H_2O}$ = (0.56$\pm$0.17)\,log\,$N_{\rm H}$ + (--11.89$\pm$3.93), with a correlation 
coefficient $r$=0.49. A trend of rising column density with maser luminosity may be apparent, but the scatter is large and the 
correlation is weak. The megamaser subsample (not displayed) shows no trend at all. There is no exponential growth of intensity 
with column density as expected in the case of unsaturated maser emission (e.g., Goldreich et al. 1972). A correlation roughly 
fitting $L_{\rm H_2O}$ $\propto$ $N_{\rm H}^{3}$ would be expected in the case of idealized saturated maser emission (no velocity 
gradients in the maser region) with the value of the exponent being determined by the luminosity increasing linearly with column 
density and the surface of the masing cone growing with the square of its lengths (e.g., Kylafis \& Norman 1991). As already indicated, 
even slight differences in the lines of sight as well as amplification by a background continuum source may be responsible for the 
weakness of the correlation.

\section{Conclusions}

In this paper, maser emission and column density are discussed for 64 extragalactic H$_2$O maser sources. The main results are:
 
(1) In searching for the $\lambda$$\sim$1.3\,cm (22\,GHz) water vapor line in galaxies with high nuclear column densities, 
active nuclei or high infrared luminosities, we detected H$_2$O emission towards NGC\,2273, UGC\,5101, and NGC\,3393 with 
isotropic luminosities of 7, 1500, and 400\,L$_{\odot}$, respectively. UGC\,5101, one of the most luminous H$_2$O megamasers 
ever found, is the first luminous ($L_{\rm H_2O}$$>$100\,L$_{\odot}$) H$_2$O maser detected in an ULIRG. The maser is 
likely nuclear. NGC\,2273 hosts a maser that is either nuclear or associated with massive star formation. Since its lineshape 
is wide, suggesting a long lifetime, interferometric measurements may soon reveal its nature. NGC\,3393, first detected by 
Kondratko et al. (2006), shows a classical NGC\,4258-like profile. If the analogy holds, the rotation velocity of the disk 
is well below the 1000\,km\,s$^{-1}$ encountered in NGC\,4258. This and similarly low rotation velocities in other `disk-maser' 
galaxies suggest that the accretion disk in NGC\,4258, surrounding a nuclear engine of relatively high mass, is particularly compact 
and well ordered.

(2) {\it Chandra} X-ray images and spectra are presented and analyzed for NGC\,2782 and NGC\,5728. A hidden AGN was detected 
in NGC\,2782. The X-ray spectrum is best fit by a high energy reflection model with $N_{\rm H}$$\ga$10$^{24}$cm$^{-2}$; its 
6.4\,keV Fe line appears to be buried in a spatially more extended, relatively bright continuum. The H$_2$O maser, so 
far believed to be associated with star formation, may be related to the nuclear engine. The X-ray spatial analysis of NGC\,5728 
confirms the biconical axisymmetric structure of its ionization region, which was previously revealed by optical HST emission line 
studies. The best fitting model, with a partially absorbed power law spectrum, yields $N_{\rm H}$$\sim$8$\times$10$^{23}$\,cm$^{-2}$. 
 
(3) A statistical analysis of the correlation between X-ray absorbing column density and H$_2$O emission was carried out, 
encompassing 64 galaxies beyond the Magellanic Clouds. At a confidence level of at least 99.4\%, the column density 
distributions of kilomaser and megamaser sources are different. More than 80\% (22/26) of the megamaser sources are heavily 
obscured ($N_{\rm H}$$>$10$^{23}$cm$^{-2}$) and half are Compton-thick ($N_{\rm H}$ $\geq$10$^{24}$cm$^{-2}$). Among the 13 
known kilomaser sources ($L_{\rm H_2O}$ $<$ 10\,L$_{\odot}$), however, only NGC\,2273 and NGC\,5194 are Compton-thick. The 
small subsamples of nuclear kilomasers and disk megamasers show average column densities that are undistinguishable from those 
of the entire megamaser sample and from samples of Seyfert 2 galaxies that were not selected with respect to maser emission. 
Presumably a clumpy cloud structure in the cirumnuclear environment, diverging positions between maser and nuclear sources and 
an occasional amplification of a background radio continuum source are sufficiently decoupling X-ray column densities and H$_2$O 
maser properties not to show a clear correlation.

(4) The identification of an AGN of type 2 is sufficient to justify a search for H$_2$O maser emission. The determination of 
X-ray column densities toward the nuclear engine does not enhance maser detection probabilities.

\begin{acknowledgements}

We wish to thank M. Elitzur for useful discussions and an anonymous referee for critically reading the text. This research 
was supported in part by NASA LTSA grant NAG 513065 to the University of Maryland and has made use of the NASA/IPAC Extragalactic 
Database (NED) which is operated by the Jet Propulsion Laboratory, CalTech, under contract with NASA. We also profited from NASA's 
Astrophysics Data System Abstract Service.

\end{acknowledgements}

\clearpage

\begin{longtable}{llclcl}
 \caption{\label{coldensity}Nuclear X-ray absorbing column densities of extragalactic H$_{2}$O maser galaxies$^{*}$}\\
\hline
 Source &  Type$^{\rm a}$ & Telescope$^{\rm b}$ &  Epoch$^{\rm c}$   & $N_{\rm H}$            & References$^{\rm d}$ \\
        &                 &                     &                    & (10$^{23}$\,cm$^{-2}$) &                      \\
\hline
\endfirsthead
\caption{continued.}\\
\hline
 Source &  Type$^{\rm a}$ & Telescope$^{\rm b}$ &  Epoch$^{\rm c}$   & $N_{\rm H}$            & References$^{\rm d}$ \\
        &                 &                     &                    & (10$^{23}$\,cm$^{-2}$) &                      \\
\hline
\endhead
\hline
\endfoot

{\it IC\,10}                    & X-1               & C & 2003.03.12 & $ 0.06^{+0.002}_{-0.0008}$ & BB04               \\
NGC\,235A                       & S2                & R & \nodata    & see Table 3                & HEASARC            \\ 
{\it NGC\,253}                  & SBG               & B & 1996.11.29 & $ 0.12_{-0.04}^{+0.03}$    & Cap99b             \\ 
                                &                   & C & 1999.12.16 & $ 2.0^{+1.3}_{-0.9}$       & Wea02              \\
NGC\,262 (Mrk\,348)             & S2                & A & 1995       & $1.60_{-0.10}^{+0.20}$     & Ris02              \\
IRAS F01063--8034$^{\rm e)}$    & S2                & X & 2003.04.21 & see Sect.\,4.2             &                    \\
NGC\,449 (Mrk\,1)               & S2                & A & 1996.08.06 & see Table 3                & TARTARUS           \\
{\it NGC\,598 (M\,33)}          &                   & A & 1993.07    & $0.016^{+0.002}_{-0.002}$  & Tak94              \\
                                & X-8               & X & 2000--2002 & $0.019^{+0.005}_{-0.005}$  & Fos04              \\
NGC\,591                        & S2                & A & 1997.07.28 & see Table 3                & TARTARUS           \\
NGC\,613                        & S                 & R & \nodata    & see Table 3                & HEASARC            \\ 
IC\,184                         & S2                & R & \nodata    & see Table 3                & HEASARC            \\ 
NGC\,1052                       & LINER             & A & 1996.08    & $ 2.0_{-0.83}^{+0.79}$     & Ter02              \\
                                &                   & X & 2001.08.15 & $\sim7$                    & Kad06              \\
NGC\,1068                       & S2, C             & A & 1993.07.24 & $>100$                     & B99, Iwa97         \\ 
                                &                   & B & 1996, 1998 & $>100$                     & Mat97, Gua99       \\ 
                                &                   & C & 2000.02.21 & $>100$                     & Mat04              \\ 
                                &                   & X & 2000.07.29 & $>100$                     & Mat04              \\ 
Mrk\,1066                       & S2, C             & A & 1997.07.24 & $>10$                      & Ris99              \\
NGC\,1386                       & S2, C             & B & 1996.12.10 & $>10$                      & Mai98              \\ 
                                &                   & X & 2002.12.29 & $\geq 22$                  & Gua05              \\ 
{\it IC\,342}                   &                   & A & 2000.02.24 & $0.084^{+0.042}_{-0.042}$  & Kub02              \\
                                & X-21              & X & 2001.02.11 & $0.087^{+0.013}_{-0.025}$  & Kon03              \\
UGC\,3255                       & S2                & R & \nodata    & see Table 3                & HEASARC            \\
Mrk\,3                          & S2, C             & A & 1993       & $13^{+22}_{-6}$            & Tur97b             \\
                                &                   & B & 1997.04.16 & $12.7^{+2.4}_{-2.2}$       & Cap99a, Matt00     \\
                                &                   & X & 2000.10.19 & $13.6^{+0.3}_{-0.4}$       & Bia05              \\
{\it NGC\,2146}                 & SBG               & A & 1997.03.26 & $0.021^{+0.022}_{-0.011}$  & Del99              \\
VII\,ZW 73                      & S2                & R & \nodata    & see Table 3                & HEASARC            \\
{\it NGC\,2273}                 & S2, C             & B & 1997.02.12 & $>100$                     & Mai98              \\
                                &                   & X & 2003.09.05 & $\geq 18$                  & Gua05              \\  
Mrk\,78                         & S2                & R & \nodata    & see Table 3                & HEASARC            \\
Mrk\,1210                       & S2, C?            & A & 1995       & $>10$                      & B99                \\
                                &                   & B & 2001.05.05 & $1.83^{+0.08}_{-0.12}$     & Ohno04             \\
                                &                   & X & 2001.05.05 & $2.14^{+0.20}_{-0.16}$     & Gua02b             \\
J0804+3607                      & Type 2 Quasar     & R & \nodata    & see Table 3                & HEASARC            \\     
NGC\,2639                       & LINER             & A & 1997.04.16 & $4.2^{+5.6}_{-2.3}$        & B99, Wil98         \\
NGC\,2782$^{\rm e)}$            & SBG, C            & C & 2002.05.17 & $>10$                      & this paper         \\
NGC\,2824 (Mrk\,394)            & S2                & R & \nodata    & see Table 3                & HEASARC            \\
NGC\,2960 (Mrk\,1419)           & LINER             & R & \nodata    & see Table 3                & HEASARC            \\
UGC\,5101                       & ULIRG             & C & 2001.05.28 & 0.1 ($<$0.65)              & Ptak03             \\
                                &                   & C,X&2001       & $13^{+2}_{-2}$             & Arm04              \\
NGC\,2979                       & S2                & R & \nodata    & see Table 3                & HEASARC            \\
{\it NGC\,3034 (M\,82)$^{\rm f)}$}& SBG             & B & 1997.12.06 & $0.058^{+0.014}_{-0.015}$  & Cap99b             \\
                                &                   & X & 2001.05.06 & $0.017^{+0.018}_{-0.016}$  & Ste03              \\ 
                                &                   & C & 2002       & 0.012                      & Str04              \\ 
NGC\,3079                       & S2/LINER, C?      & A & 1993.05.09 & $0.17^{+0.18}_{-0.12}$     & B99, Ter02         \\
                                &                   & B & 2000.05.26 & $\sim 100$                 & Iyo01              \\ 
                                &                   & C & 2001.03.07 & $0.17^{+0.02}_{-0.02}$     & Cec02              \\   
Mrk\,34                         & S2                & A & 1997.11.17 & see Table 3                & TARTARUS           \\
IC\,2560                        & S2                & A & 1996.12.19 & $\sim$3                    & Ish01              \\
                                &                   & C & 2000.10.29 & $>10$                      & Iwa02              \\
                                &                   &C  & 2004.02.16 & $\ga$30                    & Mad06              \\
NGC3393                         & S2, C             & X & 2003.01.05 & $44^{+25}_{-11}$           & Gua05              \\
{\it NGC3556}                   & X-35              & C & 2001.09.08 & $0.03^{+0.017}_{-0.009}$   & Wang03             \\   
Arp\,299 (NGC\,3690)            & SBG, C            & X & 2001.05.06 & $\sim 26$                  & Ball04             \\
                                &                   & C & 2001.07.13 & $25^{+4.4}_{-2.5}$         & Zez03              \\  
                                &                   & B & 2001.12.14 & $25.2^{+1.39}_{-0.56}$     & Del02              \\ 
NGC\,3735                       & S2                & R & \nodata    & see Table 3                & HEASARC            \\
{\it NGC\,4051}                 & S1/1.5            & A & 1993.04.25 & $0.045_{-0.035}^{+0.035}$  & Geo98              \\
                                &                   & C & 2000.04.25 & $\sim 0.01$                & {\rm Col01}        \\
                                &                   & X & 2001.05.16 & $\sim 3.6$                 & Pou04              \\
{\it NGC\,4151}                 & S1.5              & A & 1993.12.07 & $0.635_{-0.035}^{+0.041}$  & Geo98              \\
                                &                   & A,B& 1999-2000 & $2.0^{+0.1}_{-0.1}$        & Sch02a             \\
NGC\,4258                       & S1.9              & A & 1999.05.15 & $0.95^{+0.21}_{-0.09}$     & Rey00              \\ 
                                &                   & B & 1998.12.19 & $1.21^{+0.07}_{-0.06}$     & Ris02              \\
                                &                   & C & 2000--2001 & $\sim 0.7$                 & You04              \\ 
                                &                   & X & 2000.12.08 & $0.80^{+0.04}_{-0.04}$     & Pie02              \\
                                &                   &X,C& 2000-2002  & $\sim 1.0$                 & Fru05              \\
{\it NGC\,4293}                 & LINER             & R & \nodata    & see Table 3                & HEASARC            \\ 
NGC\,4388                       & S2                & A & 1995.07.21 & $3.34^{+1.00}_{-0.90}$     & For99              \\ 
                                &                   & B & 2000.01.03 & $4.80^{+1.80}_{-0.80}$     & Ris02              \\
                                &                   & X & 2002.12.12 & $2.79^{+0.07}_{-0.07}$     & Bec04              \\ 
                                &                   & C & 2001.06.08 & $3.5^{+0.4}_{-0.3}$        & Iwa03              \\
ESO\,269-G012$^{\rm e)}$        & S2                & C & 2004.01.05 & see Sect.\,4.2             &                    \\
NGC\,4922$^{\rm g)}$            & S2                & C & 2004.11.02 & see Sect.\,4.2             &                    \\
NGC\,4945                       & S2, C             & A & 1993.08.31 & $40^{+2.0}_{-1.2}$         & B99, Don96         \\
                                &                   & B & 1999.07.01 & $22^{+3}_{-4}$             & Gua00              \\
{\it NGC\,5194 (M\,51)}         & S2, C             & A & 1993.05.11 & $7.5^{+2.5}_{-2.5}$        & B99,Ter98          \\
                                &                   & B & 2000.01.18 & $56^{+40}_{-16}$           & Fuk01              \\
NGC\,5256 (Mrk\,266)            & S2                & A & 1999.05    & $0.16^{+0.33}_{-0.15}$     & Lev01              \\
NGC\,5347                       & S2, C             & A & 1997       & $>10$                      & Ris99              \\
NGC\,5495                       & S2                & R & \nodata    & see Table 3                & HEASARC            \\
Circinus                        & S2, C             & A & 1995.08.14 & $>10$                      & B99, Mat96         \\
                                &                   & B & 1998.03.13 & $43^{+4}_{-7}$             & Mat99              \\
                                &                   & C & 2000.06.06 & $\sim 60$                  & Sam01              \\
NGC\,5506 (Mrk1376)             & S1?               & B & 1998.01.14 & $0.37^{+0.05}_{-0.05}$     & Ris02              \\
NGC\,5643                       & S2, C             & B & 1997.03.01 & $>100$                     & Mai98, B99         \\
                                &                   & X & 2003.02.08 & 6$-$10                     & Gua04              \\  
NGC\,5728$^{\rm e)}$            & S2                & C & 2003.05.27 & $7.7^{+0.55}_{-0.51}$      & this paper         \\
NGC\,5793                       & S2                & A & 1998.08.05 & see Table 3                & TARTARUS           \\
NGC\,6240                       & ULIRG, C          & A & 1994.03.27 & $>$20                      & Iwa98              \\
                                &                   & B & 1998.08.15 & $21.8^{+4.0}_{-2.7}$         & Vig99            \\ 
                                &                   & X & 2000.09.22 & $10.0^{+3.0}_{-3.0}$       & Bol03              \\
                                &                   & X & 2001.07.29 & $13.0^{+17.0}_{-8.0}$      & Ptak03             \\   
{\it NGC\,6300}                 & S2, C?            & RX& 1997.02    & $>100$                     & Lei99              \\
                                &                   & B & 1999.08    & $2.10^{+0.10}_{-0.10}$     & Gua02a             \\
                                &                   & X & 2001.03.02 & $2.15^{+0.08}_{-0.09}$     & Mats04             \\ 
NGC\,6323                       & S2                & R & \nodata    & see Table 3                & HEASARC            \\  
ESO\,103-G035                   & S2                & A & 1996.03.18 & $2.16^{+0.26}_{-0.25}$     & For99              \\
                                &                   & B & 1997.10.14 & $2.02^{+0.28}_{-0.28}$     & Wil01              \\
IRAS F19370--0131               & S2                & R & \nodata    & see Table 3                & HEASARC            \\
3C\,403                         & FRII              & C & 2002.12.07 & $4.00^{+0.20}_{-0.20}$     & Kra05              \\
NGC\,6926                       & S2                & R & \nodata    & see Table 3                & HEASARC            \\
AM2158-380NED02                 & S2                & R & \nodata    & see Table 3                & HEASARC            \\
TXS\,2226-184                   & LINER             & A & 1997.11.21 & see Table 3                & TARTARUS           \\   
IC\,1481                        & LINER             & A & 1997.11.28 & see Table 3                & TARTARUS           \\

\end{longtable}
$^{*}$ {\it Chandra} and {\it XMM-Newton} results are given whenever possible. For sources not observed by these satellites
most recent (but usually older) results  are quoted. Sources in italics: kilomaser sources.
\ \ \ \ \

{\em a)} Type of nuclear activity. In the absence of a starburst and a prominent nuclear source, the X-ray source (e.g. X-1 or X-8) is 
given. SBG: StarBurst Galaxy; S2: Seyfert 2; LINER: Low-Ionization Nuclear Emission Line Region; ULIRG: UltraLuminous InfraRed 
Galaxy; FRII: Fanarov-Riley Type II galaxy; C: Compton-thick, i.e. $N_{H}$$\ga$10$^{24}cm^{-2}$; C?: {possibly varying} between 
Compton-thick and Compton-thin

{\em b)} A: {\it ASCA}; B: {\it BeppoSax}; C: {\it Chandra}; R: {\it ROSAT}; RX: {\it RXTE}; X: {\it XMM-Newton} 

{\em c)} Year, month and day

{\em d)} 
Arm04: Armus et al. (2004);
Ball04: Ballo et al. (2004); 
B99: Bassani et al. (1999); 
BB04: Bauer \& Brandt (2004); 
Bec04: Beckmann et al. (2004);
Bia03: Bianchi et al. (2003);
Bia05: Bianchi et al. (2005);
Bol03: Boller et al. (2003);
Cap99a: Cappi et al. (1999);
Cap99b: Cappi et al. (1999); 
Cec02: Cecil et al. (2002);
Col01: Collinge et al. (2001);
Del99: Della Ceca et al. (1999);
Del02: Della Ceca et al. (2002); 
Don96: Done et al. (1996);
For99: Forster et al. (1999);
Fos04: Foschini et al. (2004);
Fru05: Fruscione et al. (2005)
Fuk01: Fukazawa et al. (2001);
Geo98: George et al. (1998); 
Gua99: Guainazzi et al. (1999); 
Gua00: Guainazzi et al. (2000);
Gua02a: Guainazzi (2002); 
Gua02b: Guainazzi et al. (2002); 
Gua04: Guainazzi et al. (2004b); 
Gua05: Guainazzi et al. (2005a);
Ish01: Ishihara et al. (2001);
Iwa97: Iwasawa et al. (1997);
Iwa98: Iwasawa \& Comastri (1998);
Iwa02: Iwasawa et al. (2002);
Iwa03: Iwasawa et al. (2003);
Iyo01: Iyomoto et al. (2001); 
Kad06: Kadler et al. (2006), in prep.;
Kon03: Kong (2003); 
Kra05: Kraft et al. (2005);
Kub02: Kubota et al. (2002)
Lei99: Leighly et al. (1999);
Lev01: Levenson et al. (2001);
Mad06: Madejski et al. (2006);
Mai98: Maiolino et al. (1998);
Mat96: Matt et al. (1996); 
Mat97: Matt et al. (1997);
Mat99: Matt et al. (1999); 
Mat00: Matt et al. (2000); 
Mat04: Matt et al. (2004);
Mats04: Matsumoto et al. (2004);
Mil90: Miller \& Goodrich (1990);
Ohno04: Ohno et al. (2004);
Per98: Persic et al. (1998); 
Pie02: Pietsch et al. (2002);
Pou04: Pounds et al. (2004); 
Ptak03: Ptak et al. (2003);
Rey00: Reynolds et al. (2000);
Ris99: Risaliti et al. (1999); 
Ris02: Risaliti et al. (2002);
Sam01: Sambruna et al. (2001);
Sch02a: Schurch et al. (2002a); 
Sch02b: Schurch et al. (2002b); 
Smi96: Smith \& Done (1996);
Ste03: Stevens et al. (2003);
Str04: Strickland et al. (2004); 
Tak94: Takano et al. (1994)
Ter98: Terashima et al. (1998); 
Ter02: Terashima et al. (2002); 
Tur97a: Turner et al. (1997a); 
Tur97b: Turner et al. (1997b); 
Vig99: Vignati et al. (1999);
Wang03: Wang et al. (2003);
Wea02: Weaver et al. (2002);
Wil01: Wilkes et al. (2001); 
Wil98: Wilson et al. (1998);
You04: Young \& Wilson (2004);
Zez03: Zezas et al. (2003).

{\em e)} Absorbing column densities were obtained by processing {\it XMM-Newton} and {\it Chandra} Archive data (see Sect.\,4.2).

{\em f)} For NGC\,3034 (M\,82), lacking a well defined nucleus, the {\it Chandra} data refer to the diffuse
halo component.

{\em g)} {\it Chandra} archive data have not yet been released.

\end{document}